\documentclass[twocolumn,showpacs,floatfix,prb]{revtex4}
\usepackage{graphicx}
\usepackage{amsmath}
\usepackage{bm}

\begin{document}

\title{Josephson current transport through a Quantum Dot in an Aharonov-Bohm Ring}

\author {Shu-guang Cheng$^1$, Yanxia Xing$^1$, X. C. Xie$^{2,1}$, and Qing-feng Sun$^{1,\ast}$}

\address{
$^1$Institute of Physics, Chinese Academy of Sciences, Beijing
100190, China\\
$^2$Department of Physics, Oklahoma State University, Stillwater,
Oklahoma 74078}

\begin{abstract}
The Josephson current through an Aharonov-Bohm (AB) interferometer,
in which a quantum dot (QD) is situated on one arm and a magnetic
flux $\Phi$ threads through the ring, has been investigated. With
the existence of the magnetic flux, the relation of the Josephson
current and the superconductor phase is complex, and the system can
be adjusted to $\pi$ junction by either modulating the magnetic flux
or the QD's energy level $\varepsilon_d$. Due to the electron-hole
symmetry, the Josephson current $I$ has the property
$I(\varepsilon_d,\Phi)=I(-\varepsilon_d,\Phi+\pi)$. The Josephson
current exhibits a jump when a pair of Andreev bound states aligns
with the Fermi energy. The condition for the current jump is given.
In particularly, we find that the position of the current jump and
the position of the maximum value of the critical current $I_c$ are
identical. Due to the interference between the two paths, the
critical current $I_c$ versus the QD's level $\varepsilon_d$ shows a
typical Fano shape, which is similar to the Fano effect in the
corresponding normal device. But they also show some differences.
For example, the critical current never reaches zero for any
parameters, while the current in the normal device can reach zero at
the destruction point.
\end{abstract}

 \maketitle

\maketitle

\section{introduction}
Mesoscopic electron transport through an Aharonov-Bohm (AB)
interferometer has attracted considerable attention recently because
of its applications in nano-technology. It reveals information about
intrinsic quantum states by detecting interference of electrons in
different paths. Interference between a continuum energy spectrum
and a discrete energy state gives transmission probability $T(E)$
asymmetric line shape of typical Fano resonance.\cite{ref1} Proposed
first by Fano, it has been broadly studied and observed in recent
experiments.\cite{ref2,ref3} By setting a quantum dot (QD) in one
arm of the AB interferometer, the Fano resonance and Kondo-Fano
resonance are found.\cite{ref4} Interference between direct
transmission and QD makes the transport phase through QD
observable.\cite{ref5} In addition, the extra phase $\Phi$ due to
either the magnetic flux or the spin-orbital interaction can
modulate the interference.\cite{ref6,ref7} The transmission
probability $T(E)$ shows periodical function of the extra phase. In
such an AB-Fano system, the $T(E)$ versus QD level shows a
Breit-Wigner resonance when the phase $\Phi$ is $\pi/2$ and a
typical Fano resonance when the phase is $0$ or $\pi$. The $T(E)$
also shows a Breit-Wigner resonance when the direct tunneling is
broken and a Fano type when direct tunneling is increased. Both the
phase parameter and direct tunneling can be included in a Fano
parameter $q$ with the transmission probability $
T(E)\propto\frac{(\epsilon+q)^2}{\epsilon^2+1}$, in which
$\epsilon=\frac{E-\varepsilon_d}{\Gamma}$, and $E$ is the energy of
the incident electron, $\varepsilon_d$ is the QD's level and
$\Gamma$ is the coupling between QD and the lead. The Fano parameter
$q$, generally a complex number, decides the line shape of the
resonance.

Previous work on the AB-Fano interferometer focused mainly on the
normal device, in which the two external leads are normal. By
attaching the interferometer to two superconductor leads instead of
two normal leads, the Cooper pair transport and Andreev tunneling
occur, then Josephson current emerges even at the zero bias. The
purpose of this paper is to study how the Josephson current is
affected by the interference of the two paths, and whether the
Josephson current also has the Fano characters as in normal systems.
Recently, the Josephson current through a mesoscopic system has been
extensively investigated because of scientific interest and possible
applications. The Josephson current through a clean thin
superconductor-normal-superconductor (S-N-S) junction has a
discontinuous jump at superconductor phase difference $\theta=\pi$
under proper conditions.\cite{ref8} This current jump arises from
the discontinuity of supercurrent contribution of Andreev
bound-states driven by phase difference $\theta$ of coupled
leads.\cite{ref9} In this situation, the current jump can be broken
by impurities,\cite{ref10} the finite temperature, the normal lead
attachment,\cite{ref11} or the electron-electron
interaction.\cite{ref12} Another interesting characteristic of the
Josephson current is the $\pi$ junction, where the sign of Josephson
current can be reversed from $I_c\sin(\theta)$ to
$I_c\sin(\theta+\pi)$.\cite{ref13} The $\pi$ junction is broadly
researched in S-ferromagnetic-S (S-F-S) junctions,\cite{ref14,ref15}
S-F ring\cite{ref16}, S-QD-S junctions,\cite{ref17,ref18} S-AB-S
junctions,\cite{ref19,ref20} and so on.\cite{ref21}

In this paper, we will investigate the Josephson current $I$
through an AB-Fano interferometer consisting of a QD in one arm
and a magnetic flux $\Phi$ through the ring. By using the
non-equilibrium Green's function method, the Josephson current
expression is obtained. Due to the electron-hole symmetry, the
Josephson current has the property
$I(\varepsilon_d,\Phi)=I(-\varepsilon_d,\Phi+\pi)$. Without the
magnetic flux, the current-superconducting phase ($I$-$\theta$)
relation usually shows a sinusoidal line shape. But when the
magnetic flux exists, the curve of $I$-$\theta$ is quite complex
and can be modulated to $\pi$ junction in appropriate parameters.
The system has two pairs of Andreev bound states due to the direct
arm and the QD. When one pair of Andreev bound states is in a line
with the Fermi energy $E_F=0$, the Josephson current jumps. The
conditions for this current jump are given. In particular, we find
that the position of the current jump is identical to the position
of the maximum value of the critical current $I_c$. The critical
current $I_c$ versus the QD level $\varepsilon_d$ shows a typical
Fano shape due to the interference between the two paths. The
positions for the constructive and destructive interferences are
same with the corresponding normal device. But the critical
current can not reach zero at the destructing position, which is
different from the normal device in which the current can be zero
at the destructing position when the magnetic flux $\Phi=n\pi$
(integer n). In addition, the critical current is a periodic
function of the magnetic flux with the period $\pi$ at the QD
level $\varepsilon_d=0$ and period $2\pi$ while
$\varepsilon_d\neq0$.

The rest of the paper is organized as follows. In Sec. II, the
Hamiltonian is present and the Josephson current expression is
derived. Main numerical results are given in Sec. III, in which we
investigate the Josephson current-superconducting phase relation,
the condition of the jump of the Josephson current, and the
characters of the critical current. A brief conclusion is given in
Sec. IV.

\section{model and formulation}
The system we considered is the AB-Fano interferometer consisting of
a QD and a reference arm connected to two BCS superconductor leads.
The Hamiltonian of the superconductor leads is
$H_{\alpha}=\sum_{k\sigma}\varepsilon_k C_{k\sigma,\alpha}^\dagger
C_{k\sigma,\alpha}+\sum_k (\Delta_\alpha C_{k\downarrow,\alpha}
C_{-k\uparrow,\alpha}+\Delta^*_\alpha C_{-k\uparrow,\alpha}^\dagger
C_{k\downarrow,\alpha}^\dagger)$, where $\alpha=L,R$ represent the
left and right lead, and $\Delta_\alpha=\Delta e^{i\theta_\alpha}$
is the complex superconducting order parameter, with the
superconductor gap $\Delta$ and the superconductor phase
$\theta_\alpha$. Coupling Hamiltonian between leads and QD is,
$H_{T}=\sum_{k\sigma,\alpha}(t_\alpha C_{k\sigma,\alpha}^\dagger
d_\sigma+t_\alpha d_\sigma^\dagger
C_{k\sigma,\alpha})+\sum_{k,k',\sigma}(t_{RL}C_{k\sigma,L}^\dagger
C_{k'\sigma,R}+t_{LR}C_{k'\sigma,R}^\dagger C_{k\sigma,L})$.
Parameter $t_{\alpha}$ is the coupling coefficient of the QD to
$\alpha$-th lead, and
$t_{LR}=t^*_{RL}=te^{i\Phi}=te^{i2\pi\frac{\phi}{\phi_0}}$ is the
direct coupling between the two leads. With the consideration of the
magnetic flux $\phi$ through AB interferometer, a phase $\Phi = 2\pi
\phi/\phi_0$ is added to the coupling coefficient $t$, with
$\phi_0=e/h$ the flux quantum. We adopt the single level QD with
neglecting the intra-dot Coulomb repulsion; its Hamiltonian is
$H_{dot}=\sum_{\sigma} \varepsilon_d d^\dagger_\sigma d_\sigma$. The
total system Hamiltonian is $H=H_L+H_R+H_{dot}+H_T$.

For the sake of calculation, we take an unitary transformation with
$U=\exp\{\sum_{k\sigma,\alpha}\frac{i\theta_\alpha}{2}
C_{k\sigma,\alpha}^\dagger C_{k\sigma,\alpha}\}$,\cite{ref22} and
the Hamiltonians are transformed to the following forms:
\begin{eqnarray}
H_{\alpha}&=&\sum_{k\sigma}\varepsilon_k
C_{k\sigma,\alpha}^\dagger C_{k\sigma,\alpha}\nonumber \\
&+&\sum_k \Delta ( C_{k\downarrow,\alpha}
C_{-k\uparrow,\alpha}+C_{-k\uparrow,\alpha}^\dagger
C_{k\downarrow,\alpha}^\dagger)
\nonumber\\H_T&=&\sum_{k\sigma,\alpha}(t_\alpha
e^{\frac{i\theta_\alpha}{2}} C_{k\sigma,\alpha}^\dagger
d_\sigma+t_\alpha e^{\frac{-i\theta_\alpha}{2}}
d_\sigma^\dagger C_{k\sigma,\alpha})\nonumber \\
&+&\sum_{k,k',\sigma}(te^{-i \frac
{\varphi}{2}}C_{k\sigma,L}^\dagger C_{k'\sigma,R}+te^{i \frac
{\varphi}{2}} C_{k'\sigma,R}^\dagger
C_{k\sigma,L})\nonumber \\
H_{dot}&=& \sum_{\sigma}\varepsilon_d d^\dagger_\sigma d_\sigma.
\end{eqnarray}
with $\varphi=2\Phi-\theta_L+\theta_R$.

The current $I$ through AB-Fano interferometer can be calculated
from the evolution of the electron number operator $N_L=
 \sum_{k\sigma}C_{k\sigma L}^\dagger C_{k\sigma L} $ in the left
leads,\cite{ref23,ref24}
\begin{eqnarray}
I&=&-e\langle \dot{N}\rangle  \\&=&\frac{4e}{\hbar} \mathrm{Re} \int
\frac{dE}{2 \pi} [t_Le^{\frac{i\theta_L}{2}}G^<_{dL,11}(E)+te^{-i
\frac {\varphi}{2}}G^<_{RL,11}(E)] \nonumber
\end{eqnarray}
Here the Nambu representation has been used. The Green's function
${\bf G}^<_{dL}(E)$ and ${\bf G}^<_{RL}(E)$ are the Fourier
transformation of ${\bf G}^<_{dL}(t-t')$ and ${\bf
G}^<_{RL}(t-t')$, which are defined as:
\begin{eqnarray}
&& {\bf G}^<_{d,L}(t-t')  \nonumber \\
&& = i\sum_k\left(
                                               \begin{array}{cc}
                                                 \langle
C_{k\uparrow,L}^\dagger(t')d_\uparrow(t) \rangle & \langle
C_{k\uparrow,L}^\dagger(t')d_\downarrow^\dagger(t) \rangle \\
                                                \langle
C_{-k\downarrow,L}(t')d_\uparrow(t) \rangle & \langle
C_{-k\downarrow,L}(t')d_\downarrow^\dagger(t) \rangle \\
                                               \end{array}
                                             \right)\nonumber
 \\
&& {\bf G}^<_{R,L}(t-t') \nonumber \\ &&=i\sum_{k,k'}\left(
\begin{array}{cc}
\langle C_{k\uparrow,L}^\dagger(t') C_{k'\uparrow,R}(t) \rangle &
\langle C_{k\uparrow,L}^\dagger(t') C_{-k'\downarrow,R}^\dagger(t) \rangle \\
\langle C_{-k\downarrow,L}(t') C_{k'\uparrow,R}(t) \rangle &
\langle C_{-k\downarrow,L}(t') C_{-k'\downarrow,R}^\dagger(t) \rangle \\
\end{array}
\right)\nonumber
\end{eqnarray}
Notice here that all of the Green's functions are functions of
time difference $t-t'$, because that we investigate the dc
Josephson current in the zero bias case.

Also, because of the zero bias case, the system is in equilibrium
and the fluctuation-dissipation theorem holds, we now have ${\bf
G}^<_{dL}(E)=-f(E)({\bf G}^r_{dL}(E)-{\bf G}^a_{dL}(E))$ and ${\bf
G}^<_{RL}(E)=-f(E)({\bf G}^r_{RL}(E)-{\bf G}^a_{RL}(E))$, where
$f(E) = 1/\{\exp[(E-E_F)/k_B T]+1\}$ is the Fermi-Dirac
distribution. The current expression becomes
\begin{eqnarray} \hspace{-3mm}
I=\frac{-4e}{\hbar} \int \frac{dE}{2 \pi}f(E) \mathrm{Re}[({\bf
G}^r_{dL}-{\bf G}^a_{dL})t_L+t^*({\bf G}^r_{RL}-{\bf
G}^a_{RL})]_{11}.
\end{eqnarray}

In the following, we need to solve the retarded Green's function
${\bf G}^r_{dL}(E)$, ${\bf G}^r_{RL}(E)$. In the Nambu
representation, the retarded Green's function of isolated
superconducting leads and QD are respectively:\cite{ref25} ${\bf
g}_{\alpha}^r(E)=-\pi\rho(E)\left(
\begin{array}{cc}
\beta(E) & \beta_0(E) \\
\beta_0(E) & \beta(E) \\
\end{array}
\right)$ and ${\bf g}_{dd}^r(E)= \left(
\begin{array}{cc}
1/(E-\varepsilon_d+i\eta) & 0 \\
0 & 1/(E+\varepsilon_d+i\eta) \\
\end{array}
\right)$, where $\rho(E)$ is the normal density of states, $\eta$ is
an infinitesimal real number, $\beta_0(E)=\beta\Delta/E$, and
$\beta(E)=E/\sqrt{\Delta^2-E^2}$ while $|E|<\Delta$ and $\beta(E)
=i|E|/\sqrt{E^2-\Delta^2}$ while $|E|>\Delta$. Tunneling
coefficients in the Nambu representation can also be expressed in
$2\times 2$ matrix with: $\textbf{t}_\alpha=t_\alpha \left(
\begin{array}{cc}
e^{i\theta_\alpha /2} & 0 \\
0 & -e^{-i\theta_\alpha /2} \\
\end{array}
\right)$ and $\textbf{t}_{LR}=t\left(
\begin{array}{cc}
e^{i\varphi /2} & 0 \\
0 & -e^{-i\varphi/2} \\
\end{array}
\right)$. In the following calculation, we take the symmetric
barriers with $t_L=t_R$ for convenience. By using Dyson's equation,
the Green's function $ {\bf \tilde{g}}^r$ of the system decoupling
with the QD (i.e. $t_L=t_R=0$) can be deduced as: ${\bf
\tilde{g}}^r_{LL}= ({\bf g}_L^{r-1} -{\bf t}_{LR} {\bf g}^r_R {\bf
t}_{LR}^* )^{-1}$, ${\bf \tilde{g}}^r_{RR}= ({\bf g}_R^{r-1} -{\bf
t}_{LR}^* {\bf g}^r_L {\bf t}_{LR} )^{-1}$, and ${\bf
\tilde{g}}_{LR}^r = {\bf g}^r_L {\bf t}_{LR} {\bf \tilde{g}}_{RR}^r
= {\bf \tilde{g}}^r_{LL} {\bf t}_{LR} {\bf g}_{R}^r$. Then the
retarded Green's function of the whole AB-Fano interferometer device
are solved as: ${\bf G}_{dL}^r={\bf G}_{dd}^r(\textbf{t}_L^*{\bf
\tilde{g}}^r_{L}+\textbf{t}_R^*{\bf \tilde{g}}^r_{RL})$, $ {\bf
G}_{RL}^r= {\bf \tilde{g}}_{RL}^r+ ({\bf
\tilde{g}}^r_{R}\textbf{t}_R+{\bf \tilde{g}}^r_{RL}\textbf{t}_L)
{\bf G}_{dd}^r(\textbf{t}_L^* {\bf
\tilde{g}}^r_{L}+\textbf{t}_R^*{\bf \tilde{g}}^r_{RL})$, and ${\bf
G}_{dd}^r=({\bf g}^{-1}_{dd}-{\bf \Sigma}^r)^{-1}$, where the
retarded self energy ${\bf \Sigma}^r$ is: $ {\bf \Sigma}^r={\bf
t}_L^*{\bf \tilde{g}}^r_{L} {\bf t}_L+ {\bf t}_R^*{\bf
\tilde{g}}^r_{R} {\bf t}_R+ {\bf t}_L^*{\bf \tilde{g}}^r_{LR} {\bf
t}_R+ {\bf t}_R^*{\bf \tilde{g}}^r_{RL}{\bf t}_L $.

After solving the retarded Green's functions ${\bf G}^r$, the
expression of the Josephson current $I$ through AB-Fano
interferometer can be reduced as:
\begin{eqnarray}
I&=&\frac{-4e}{\hbar}\int \frac{dE}{2 \pi}f(E)\mathrm{Im}
\{\frac{2x\beta_0^2\sin\varphi}{\mathcal{D}}
\nonumber \\
&-&\frac{\Gamma\beta}{\mathrm{\Xi}} (Q_{11}A_{11}+
Q_{12}A_{12}) \\
&+&\frac{\Gamma\sqrt{x} \beta^2e^{-i\frac{\theta+\varphi}{2}}}
{\mathcal{D}\mathrm{\Xi}}[A_{11}^2Q_{11}
+A_{12}^2Q_{22}+2A_{11}A_{12}Q_{12}]\}\nonumber
\end{eqnarray}
where $ \mathcal{D}=1+x^2+2x
(E^2-\Delta^2\cos\varphi)/(E^2-\Delta^2) $ and
\begin{eqnarray}
\mathrm{\Xi}&=&[E\mathcal{D}+\Gamma\beta(1+x)]^2-
[\varepsilon_d\mathcal{D}-
\Gamma\sqrt{x}(x-\beta^2)\cos\frac{\theta+\varphi}{2} \nonumber
\\ &-&\Gamma\sqrt{x} \beta_0^2\cos\frac{\theta-\varphi}{2}]^2
-\Gamma^2\beta_0^2[\cos
\frac{\theta}{2}+x\cos(\frac{\theta}{2}+\varphi)]^2.
\end{eqnarray}
Here $x \equiv t^2\pi^2\rho^2$ dictates the tunneling through the
direct arm and $\Gamma \equiv 2\pi\rho t_{\alpha}^2$ is the
coupling strength of QD to leads. In the wide-band approximation,
$x$ and $\Gamma$ are independent with the energy $E$. The factors
$A$ and $Q$ in equation (4) are:
\[
\begin{cases}
\  A_{11}=(1+x)+\frac{\sqrt{x}}{\beta}(x+\beta^2)
e^{i\frac{\theta+\varphi}{2}}-
\frac{\sqrt{x}}{\beta}\beta_0^2e^{i\frac{\theta-\varphi}{2}} \\
\ A_{12}=-\frac{\Delta}{E}e^{i\theta/2}(1+xe^{i\varphi})+2i\sin
\frac{\varphi}{2}\frac{\Delta}{E}\beta \sqrt{x}
   \\
\ A_{21}=\frac{\Delta}{E}e^{-i\theta/2}(1+xe^{-i\varphi})-
2i\sin\frac{\varphi}{2}\frac{\Delta}{E}\beta
\sqrt{x} \\
\  A_{22}=-(1+x)+\frac{\sqrt{x}}{\beta}(x+\beta^2)
e^{-i\frac{\theta+\varphi}{2}}-\frac{\sqrt{x}}{\beta}\beta_0^2
e^{-i\frac{\theta-\varphi}{2}}  \\
\end{cases}
\]
and
\[
\begin{cases}
\  Q_{11}=E\mathcal{D}+\varepsilon_d\mathcal{D}-\Gamma\beta\mathrm{Re}(A_{22}) \\
\  Q_{12}=\Gamma\beta\mathrm{Re}(A_{12}) \\
\  Q_{21}=\Gamma\beta\mathrm{Re}(A_{12})\\
\  Q_{22}=E\mathcal{D}-\varepsilon_d\mathcal{D}+\Gamma\beta\mathrm{Re}(A_{11})  \\
\end{cases}
\]

The first term in equation (4) describes the direct tunneling
contribution, the second and third terms are transport through QD
and interference between direct arm and QD. The Josephson current in
the equation (4) can be split into two parts, the continuous part
$I_{con}$, contributed from continuous spectrum while the energy $E$
outside the superconducting gap $\Delta$, and the discrete part
$I_{dis}$ contributed by Andreev bound states while $E$ within the
gap.
\begin{figure}[htbp]
\begin{center}
\centering
\includegraphics[height=2.6in]{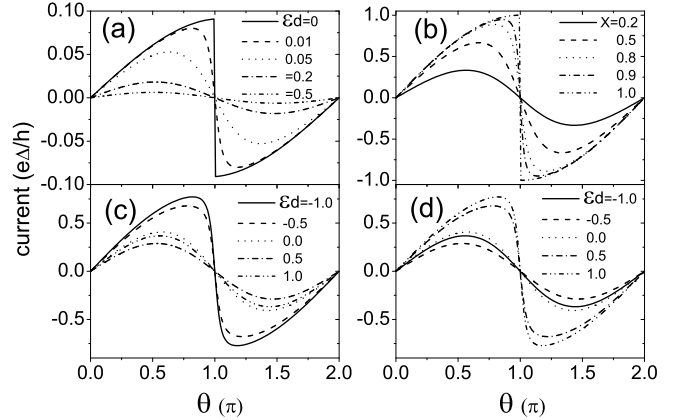}
\caption{ Current $I$ vs. superconductor phase $\theta$ for various
parameters: (a) $\Phi=0$, $x=0$, and $\Gamma=0.1$ for different QD's
level $\varepsilon_d$; (b) $\Phi=0$ and $\Gamma=0$ for different
$x$; (c) and (d) $x=\Gamma=0.5$, $\Phi=0$ (c) and $\pi$ (b) for
different $\varepsilon_d$.}
\end{center}
\end{figure}
In numerical calculation $I_{con}$ is obtained by integral in
equation (4) and $I_{dis}$ is approached by dealing with the delta
functions due to the infinitesimal imaginary part $i\eta$. The
discrete part is usually much larger than the continuous part. The
factor $\Xi$ has two pairs of poles at $E_{1,2}^{\pm}$
($E_{1,2}^{+}=-E_{1,2}^{-}$) within the gap, which count for Andreev
bound states. These Andreev bound states arise from the
hybridization of bound states of QD $E^{\pm}_{QD}$ and the direct
arm $E^{\pm}_0$. Here $E^{\pm}_0=\pm\Delta
\sqrt{1-\frac{4x}{(1+x)^2}\sin^2(\varphi/2)}$ are the Andreev bound
states in the direct arm,\cite{ref27} and $E^{\pm}_0$ are also the
poles of the factor $\mathcal{D}$. $E^{\pm}_{QD}$
($E^{+}_{QD}=-E^{-}_{QD}$) are the Andreev bound states in the QD
for the S-QD-S device.

\section{numerical results}
In this section, we present our numerical investigations on the
relation of the Josephson current versus the superconducting
phase, the condition of the jump of the current, the Fano resonant
characters of critical current, and the dependence of critical
current on the magnetic flux.

\subsection{Josephson current-superconducting phase relations}

We first discuss the current-phase ($I$-$\theta$) relation of
AB-Fano interferometer. Fig.1 shows Josephson current $I$ as a
function of the superconducting phase difference $\theta$ of the
left and right leads. When $x=0$, the device reduces into a QD
coupled to two superconductor leads. The $I$-$\theta$ relation is
sinuous line shape when the level $\varepsilon_d$ is far off the
Fermi level $E_F=0$, and the current shows a discontinuous jump at
$\theta=\pi$ when $\varepsilon_d=0$ (shown Fig.1a), in which the
Andreev bound states $E_{QD}^{+}=E_{QD}^-
=0$.\cite{ref28,ref29,ref30} On the other hand, while $\Gamma=0$,
the system reduces into an S-I-S device. The $I$-$\theta$ relation
is sinuous line shape when $x$ is much smaller or larger than 1, and
the current $I$ has a discontinuous jump while $x=1$ (see Fig.1b),
in which the Andreev bound states $E_0^+=E_0^-
=0$.\cite{ref10,ref27} When both $x$ and $\Gamma$ are non-zero (in
other words, the two pathes are opened), the AB-Fano interferometer
is formed and the transport can be adjusted by magnetic flux $\Phi$
through it. Two pairs of Andreev bound states $E_0^{\pm}$ and
$E_{QD}^{\pm}$, which belong to the direct arm and QD, cause the
hybridization to form new Andreev bound states $E_{1,2}^{\pm}$,
which enable the interference construction or destruction of the
Josephson current. Then the current-phase $I$-$\theta$ relation is
usually not a sinusoidal-like function (except for the special
magnetic flux values $\Phi =0$ and $\pi$). While $\Phi =0$ and
$\pi$, the $I$-$\theta$ relation is still a sinuous line shape, and
the current $I$ is zero at $\theta=0$ and $\pi$, as shown in Fig.1c
and 1d. In some specific parameters the discontinuous jump of the
current can still occur, which we will detail in the next
sub-section. Here, we notice that the current has the relation:
$I(\varepsilon_d, \Phi)= I(-\varepsilon_d, \Phi+\pi)$. In other
words, while the level changes from $\varepsilon_d$ to
$-\varepsilon_d$ and the magnetic flux from $\Phi$ to $\Phi+\pi$,
the current $I$ does not vary regardless of any other parameters.
The relation of $I(\varepsilon_d, \Phi)= I(-\varepsilon_d,
\Phi+\pi)$ comes from the electron-hole symmetry, i.e. taking the
transform ($d_{\sigma}, d_{k\sigma,\alpha}$) to
($\tilde{d}_{\sigma}^{\dagger},
\tilde{d}_{k\sigma,\alpha}^{\dagger}$) and simultaneously setting
the parameters ($\varepsilon_d, \Phi$) to ($-\varepsilon_d,
\Phi+\pi$), the Hamiltonian $H$ is invariable.

When the magnetic flux $\Phi$ is not equal to $0$ or $\pi$, the
current-phase relation is usually not a sinusoidal-like function,
and the current $I$ has non-zero values at $\theta=0$. In some
special parameters, the current $I$ is negative while
$\theta\in[0, \pi]$, which is a $\pi$ junction. For example, by
proper selection of parameters, $\varepsilon_d=0.5$,
$\Gamma=0.45$, $\Phi=0.6\pi$, and $x=0.4\sim0.9$, the current $I$
is negative when the phase $\theta\in[0, \pi]$ as shown in Fig.2a.
But it is not a strict $\pi$ junction, and the current is not
positive in all regions $\theta \in [\pi, 2\pi]$.\cite{ref30} The
realization of quasi-$\pi$ junction is because of the introduction
of the magnetic flux phase $\Phi$ which changes the interference
of two pathes. Another example, as shown in Fig.2b with the
magnetic flux $\Phi =\pi/2$, the current is negative (positive) in
most parts of region $\theta \in [0, \pi]$ ($[\pi, 2\pi]$). Here
we notice that the $\theta$ region for the negative current $I$ is
not equal to that of the positive current. In all curves in Fig.2a
and some curves in Fig.2b, the negative-current region is
obviously wider than the positive-current region.
\begin{figure}[htbp]
\begin{center}
\centering
\includegraphics[height=2.4in]{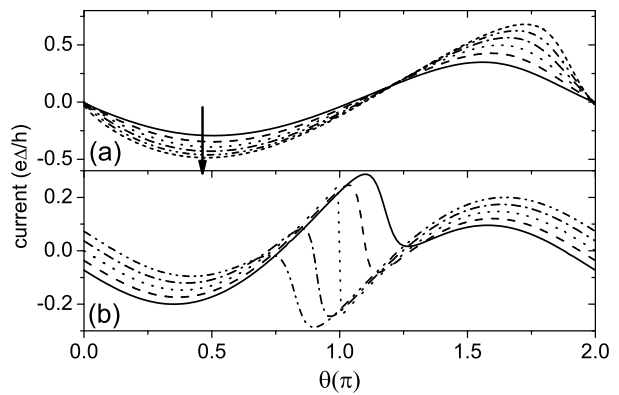}
\caption{ Current $I$ vs. the phase $\theta$ with the parameters in
(a) $\varepsilon_d$=-0.5, $\Gamma$=0.45, $\Phi$=0.6$\pi$, and $x$
from $0.4$ to $0.9$ with space 0.1 in arrow direction, (b) $x$=0.5,
$\Gamma=0.1$, $\Phi$=0, and $\varepsilon_d= -0.1$ (solid curve),
$-0.05$ (dashed curve), 0 (dotted curve), 0.05 (dash-dotted curve),
and 0.1 (dash-dot-dot curve).}
\end{center}
\end{figure}
However, no
matter the parameters, the positive-current region always exists.
In other words, it is not possible for the current to be negative
in the entire region $\theta \in [0, 2\pi]$. In addition, while
$\varepsilon_d =0$, a jump emerges in the curve of $I$-$\theta$ at
$\theta =\pi$ (see Fig.2b). We will study this discontinuous jump
in detail in the next sub-section.

\subsection{the condition of the jump of the current}

The current $I_{con}$ from continuous spectrum with $|E|>\Delta$
is always continuous. The discontinuous current arises from the
part $I_{dis}$ which is from the Andreev bound states. When one of
the two pairs of Andreev bound states $E_{1,2}^{\pm}$ just aligns
with the Fermi level $E_F=0$ (i.e. $E_{i}^+=E_{i}^-=0$), an abrupt
jump occurs in the current $I_{dis}$ so that the current
$I=I_{con}+I_{dis}$. The condition of the jump of the current $I$
is thus $ \Xi (E=0) =0$. With the help of Eq.(5), the condition $
\Xi (E=0) =0$ can be reduced to:
\begin{eqnarray}
\begin{cases}
\  \varepsilon_d\mathcal{D}- \Gamma x\sqrt{x}  \cos{\Phi}
-\Gamma\sqrt{x}\cos ({\theta-\Phi})=0 \\
\  \Gamma^2  [\cos
\frac{\theta}{2}+x\cos(2\Phi-\theta/2)] =0 \\
\end{cases}
\end{eqnarray}
\begin{figure}[htbp]
\begin{center}
\centering
\includegraphics[height=2.4in]{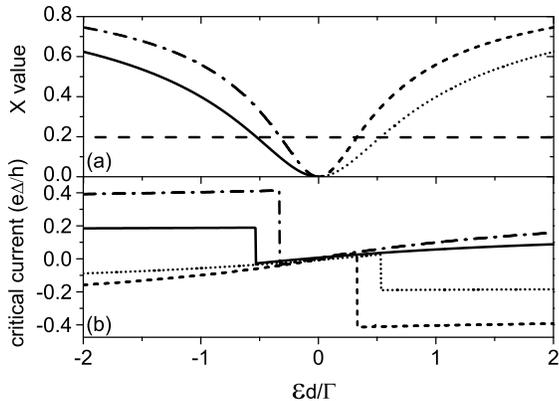}
\caption{ Upper panel shows the jump position in the parameter space
of ($x$, $\varepsilon_d/\Gamma$). The curves from left to right are
for $\Phi=0.1\pi$, $0.3\pi$, $0.7\pi$, and $0.9\pi$. Lower panel
shows the current $I$ vs. the QD's level $\varepsilon_d$ with
$x=0.2$ and $\theta=-2\arctan[(1+x\cos2\Phi)/(x\sin 2\Phi)]$. The
magnetic flux $\Phi$ for different kind of curves are same with the
upper panel. }
\end{center}
\end{figure}
When the above two equations are tenable, the current will jump.
For example, (i) while $\Gamma=0$, the condition in Eq.(6) reduces
into $x=1$ and $\theta=\pi$. This is consistent with the jump in
Fig.1b. (ii). While $x=0$, the condition in Eq.(6) reduces into
$\varepsilon_d=0$ and $\theta=\pi$, which agrees with the jump in
Fig.1a. In general, when both $x$ and $\Gamma$ are non-zero with a
magnetic flux through the AB-Fano interferometer, the condition in
Eq.(6) can be rewritten as:
\begin{eqnarray}
\theta=-2\arctan [\frac{1+x\cos2\Phi}{x\sin2\Phi}]
\end{eqnarray}
and
\begin{eqnarray}
\varepsilon_d =-\frac{\Gamma\sqrt{x}\cos\Phi}{1-x}.
\end{eqnarray}
Specifically, for $\Phi=\pi/2$, we have $\theta=\pi$ and
$\frac{\varepsilon_d}{\Gamma}= 0$, which is consistent with the
jump position in Fig.2b. Fig.3a shows the jump-occurrence region
in the parameter space of ($x$, $\frac{\varepsilon_d}{\Gamma}$).
When the parameters are just at the curves of Fig.3a, a current
jump will occur. Fig.3b shows the current $I$ versus the level
$\varepsilon_d/\Gamma$ at $x=0.2$ and $\Gamma=0.5$. It clearly
shows that the jump occurs at the corresponding intersections in
Fig.3a.

\subsection{Fano resonant characters of critical current}

In this and the next sub-section, we will focus on the critical
current, which is experimentally accessible. In fact, the critical
current in the superconducting AB-Fano interferometer behaves
similarly to the current in the normal AB-Fano interferometer
under the small bias. So here, we simply recall the results of the
normal AB-Fano device, which consists of an AB ring attached to
two normal leads with a QD in one of its arms. The transmission
probability $T$ of the normal AB-Fano device is\cite{ref4,ref6}
\begin{eqnarray}
\mathfrak{T}(E)&=&\frac{4x}{(1+x)^2}+\frac{4\Gamma(1-x)\sqrt{x}
\cos\Phi}{(1+x)^3} \mathrm{Re}G^r(E)\nonumber
\\
&-&\frac{\Gamma[(1+x)^2- 4x(1+\cos^2\Phi)
]}{(1+x)^3}\mathrm{Im}G^r(E),
\end{eqnarray}
where ${G^r}^{-1}(E)= {E-\varepsilon_d+ \Gamma\sqrt{x}\cos\Phi
/(1+x)-i\Gamma/(1+x)}$ and the meaning of the parameters
($\varepsilon_d$, $\Gamma$, $\Phi$, and $x$) is the same as the
above superconducting device. From the Eq.(9), we found that the
interference construction can be realized at
\begin{figure}[htbp]
\begin{center}
\centering
\includegraphics[height=2.4in]{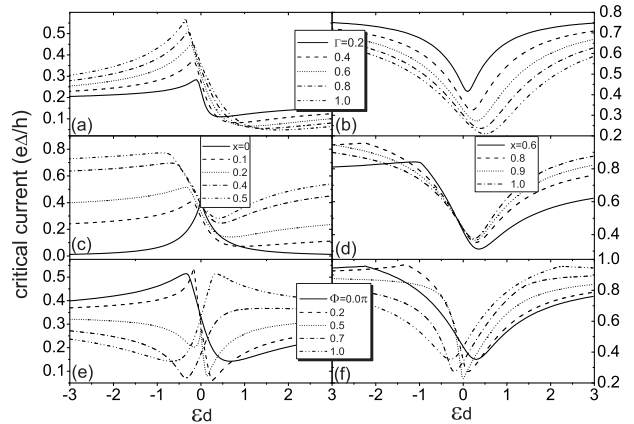}
\caption{ Critical current $I_c$ vs. $\varepsilon_d$ with parameters
in (a) $x$=0.1 and $\Phi$=0; (b) $x$=1.0 and $\Phi$=0; (c) and (d)
$\Gamma=0.5$ and $\Phi=0$; (e) $\Gamma=0.5$ and $x=0.2$; (f)
$\Gamma=0.5$ and $x=0.8$. }
\end{center}
\end{figure}
\begin{eqnarray}
\varepsilon_d=-\frac{\Gamma\sqrt{x}\cos\Phi}{1-x}
\end{eqnarray}
and interference destruction is at
\begin{equation}
\varepsilon_d=\frac{(1+x\cos2\Phi)\Gamma}{2\sqrt{x} (1+x)\cos\Phi}.
\end{equation}

Here the interference construction position is the same as Eq.(8) of
the position of the jump of the superconducting current in the
superconducting device. The $I$-$\varepsilon_d$ relation of the
normal device shows a symmetric line shape when $x=0$ and a typical
Fano resonance when at finite $x$ and zero magnetic flux $\Phi$. The
shape of the curve of $I$-$\varepsilon_d$ can be affected by the
magnetic flux $\Phi$ and is symmetric at $\Phi=\pi/2$.\cite{ref6}

Next, we study the critical current $I_c$ in the superconducting
AB-Fano interferometer. Here we select the maximum of current in a
period $2\pi$ of superconducting phase difference $\theta$ as the
critical current. Due to the interference between two pathes, the
construction or destruction transport occurs, and the critical
current $I_c$ versus the QD's level $\varepsilon_d$ usually
exhibits a Fano characters. To show the details of Fano
characters, we plot $I_c$ as a function of $\varepsilon_d$ at zero
magnetic flux ($\Phi=0$) for small $x$ values and large $x$ values
in Fig. 4(a) and (b), respectively. For small $x$ values ($x=0.1$,
for instance), $I_c$ shows a Fano resonance near $\varepsilon_d=0$
with the peak at $\varepsilon_d= -\Gamma\sqrt{x}/(1-x)$ and the
valley at approximately $\varepsilon_d= \Gamma/(2\sqrt{x})$. The
positions of the peak and valley are the same with the normal
device [see Eqs.(10) and (11)], though in the normal device the
normal current is under the small bias, while in the present
superconducting device the critical current $I_c$ is at zero bias.
With the enhancement of $\Gamma$, the line shape tends to behave
more symmetric characters as the Breit-Wigner resonance and the
magnitude of the critical current is enhanced at negative
$\varepsilon_d$ side but reduced at positive $\varepsilon_d$ side.
For large $x$ values (e.g. $x=1.0$, as shown in Fig.4b), the
destructive interference plays the core role, and the curves of
$I_c$-$\varepsilon_d$ are almost symmetric for the small $\Gamma$
and show an obvious Fano valley at $\varepsilon_d=\Gamma/2$. The
Fano peak is pushed to infinity at $x=1.0$ (as shown in Eq.(10)),
so it is invisible in Fig.4(b). With $\Gamma$ increasing, the
transmission probability through the QD grows and the destruction
of two paths occurs, causing the valley to deepen and the line
shape to become more asymmetric.

In this paragraph, we investigate the effect of the direct path
$x$ on the critical current $I_c$. In Figs. 4(c) and 4(d), we show
the critical current $I_c$ versus the QD's level $\varepsilon_d$
at different $x$ for a fixed $\Gamma =0.5$ and $\Phi=0$. While
$x=0$, the direct path is closed, $I_c$ is completely symmetric
with the peak at $\varepsilon_d=0$ and the valley at
$\varepsilon_d=\infty$, which are the same with Eqs.(10) and (11).
When $x$ increases, the peak position is moved off from the Fermi
level $E_F=0$ but the valley position gradually from infinity to
$0$, and the curve of $I_c$-$\varepsilon_d$ shows a Fano
resonance. At the middle $x$, the Fano resonance is the most
prominent. While $x$ tends to $1$, the direct path is completely
open and the peak position tends to infinity, leading the
$I_c$-$\varepsilon_d$ curve into a symmetric valley.

In the above $I_c$-$\varepsilon_d$ relation discussion, the
magnetic flux $\Phi$ is fixed at zero. In the normal AB-Fano
device, the line shape of the current $I$ versus $\varepsilon_d$
is strongly modified by the magnetic flux $\Phi$. So, in the
following, we investigate how the $I_c$-$\varepsilon_d$ curve in
the superconducting AB-Fano device is affected by $\Phi$.
Figs.4(e) and (f) show $I_c$-$ \varepsilon_d$ with $\Gamma>x$ and
$\Gamma<x$, respectively. These two cases when $\Gamma>x$ and
$\Gamma<x$ represent situations where the QD path or direct arm
path dominates the transport. While $\Gamma>x$, the
$I_c$-$\varepsilon_d$ curve is strongly affected by the magnetic
flux $\Phi$. When $\Phi$ increases from $0$, the Fano shape first
grows more notable, then the peak is greatly reduced while $\Phi$
near $\pi/2$, and at last the peak is increased and the Fano shape
is recovered again at $\Phi=\pi$ (see Fig.4e). On the other hand,
while $\Gamma<x$, the $I_c$-$\varepsilon_d$ curve is only slightly
affected by $\Phi$ (see Fig.4f). We add the following three
points: (i) The critical current $I_c$ has the relation:
$I_c(\varepsilon_d,\Phi)=I_c(-\varepsilon_d,\Phi+\pi)$, because of
the electron-hole symmetry and
$I(\varepsilon_d,\Phi)=I(-\varepsilon_d,\Phi+\pi)$. Due to the
relation $I_c(\varepsilon_d,\Phi)=I_c(-\varepsilon_d,\Phi+\pi)$,
the critical current $I_c$ is a periodic function of $\Phi$ with
period $\pi$ at $\varepsilon_d=0$. (ii) At $\Phi=\pi/2$, the
$I_c$-$ \varepsilon_d$ curve is still not symmetric, which is
different from the current in the normal device. (iii) Though the
valley in the normal device can reach zero in some parameter
regions, at no parameters does the valley value of the critical
current $I_c$ reach zero.

\begin{figure}[htbp]
\begin{center}
\centering
\includegraphics[height=2.8in]{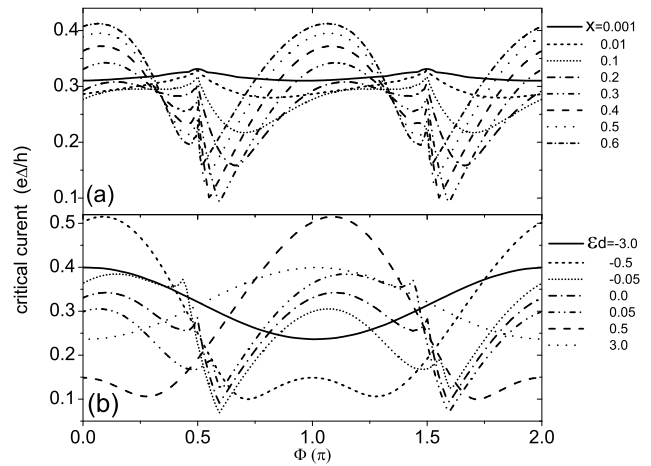}
\caption{Upper panel: critical current $I_c$ vs $\Phi$ for different
$x$ with $\varepsilon_d=0$ and $\Gamma=0.5$. Lower panel: critical
current $I_c$ vs. $\Phi$ for different $\varepsilon_d$ with $x=0.2$
and $\Gamma=0.5$.}
\end{center}
\end{figure}

\subsection{Critical current-magnetic flux relations}

Finally, we investigate the relation of the critical current $I_c$
with the magnetic flux phase $\Phi$. Fig.5(a) shows $I_c$ versus
$\Phi$ for the different $x$ values at $\varepsilon_d=0$. Several
characteristics can be noticed: i) $I_c$ is a periodic function of
$\Phi$ (or the magnetic flux $\phi$) with the period $2\pi$ (or
$e/h$) while $\varepsilon_d\not=0$, and period $\pi$ (or $e/2h$)
at $\varepsilon_d=0$. ii) For a small $x$ value, $I_c$ is almost a
constant, because the direct path is almost closed. For a large
$x$ value, the oscillation of $I_c$-$\Phi$ is strong. The
oscillation shape greatly departs from $\sin\Phi$ or $\cos\Phi$
shape because the higher order tunneling processes are numerous at
$\varepsilon_d=0$. iii) The critical current $I_c$ shows a peak at
$\Phi=(2n+1)\pi/2$ with the integer $n$, which is consistent with
equations (7) and (10). Fig.5(b) shows $I_c$ versus $\Phi$ for the
QD level $\varepsilon_d$ at $x=0.2$. When $\varepsilon_d$ is far
away from zero, $I_c$ shows a standard $\sin\Phi/\cos\Phi$
behavior because the higher order tunneling processes are weak at
the small $x$ and $|\varepsilon_d|\gg 0$. But $I_c$ for positive
$\varepsilon_d$ bears a phase lapse of $\pi$ with that of
$-\varepsilon_d$. This phase lapse is from the phase $\Theta_{QD}$
of the transmission coefficient of a QD, with $\Theta_{QD}=\pi/2$
when $\varepsilon_d\gg0$ and $-\pi/2$ when $\varepsilon_d\ll0$. As
$\varepsilon_d$ approaches zero, the oscillation of $I_c$-$\Phi$
is enhanced and departs from the $\sin\Phi/\cos\Phi$ behavior,
with peaks emerging at about $\Phi=\pi/2$ and $3\pi/2$.

\section{conclusion}

In conclusion, the Josephson current through an Aharonov-Bohm
interferometer consisting of a quantum dot and a direct arm with
magnetic flux through the ring has been investigated. An equation
for the occurrence of supercurrent discontinuity is given. In
particular, we found that the position of the supercurrent
discontinuity, the position of the peak of critical current, and
the constructive interference of the current in a corresponding
normal device are the same. By adjusting the device's parameters,
such as the magnetic flux phase and direct arm coupling, the
Josephson junction can vary from a normal junction to a
$\pi$-junction. Fano characters of the critical current are
similar to those of the current in the normal device in the small
bias, but there are also some differences. For example, the
critical current cannot reach zero with any parameters, while the
current in the normal device can reach zero in the destruction
position. Also, the critical current with variation of magnetic
flux shows a period of $e/{2h}$ when QD level aligns to the Fermi
level, or a quantum magnetic flux of $e/h$ when the QD level apart
from the Fermi level.

\section*{Acknowledgments} We gratefully acknowledge the
financial support from the Chinese Academy of Sciences and
NSF-China under Grants Nos. 10525418, 10734110, and 60776060.
X.C.X. is supported by US-DOE under Grants No. DE-FG02- 04ER46124
and NSF.

\end{document}